# KAON INTERFEROMETRY AT KLOE: PRESENT AND FUTURE

The KLOE collaboration*

Presented by A. Di Domenico#

Dipartimento di Fisica, Università di Roma "La Sapienza" and INFN sezione di Roma, Italy

*Abstract*

Interferometry with neutral kaons at KLOE is reviewed. The preliminary results on $\phi \to K_S K_L \to \pi^+\pi^- \pi^+\pi^-$ and decoherence parameter are presented. The impact of C-even background and regenerators is shortly discussed and updated. Finally KLOE perspectives on interferometry at DAΦNE and expectations at a new higher luminosity φ-factory are briefly illustrated.

## INTRODUCTION

Neutral kaon interferometry is one of the most peculiar feature of DAΦNE [1], and it will certainly be one of the most important issues at a future higher luminosity φ-factory (for the sake of clarity from now on it will be called DAΦNE-2). In the following the KLOE status and perspectives are reviewed and, in the spirit of this workshop, the potentialities offered by DAΦNE-2 will be briefly discussed.

The φ production cross section at DAΦNE is $\sigma_\phi \sim 3$ μb corresponding to $\sim 10^6$ neutral kaon pairs produced per pb$^{-1}$ of integrated luminosity. The pair is produced in antisymmetric quantum state $J^{PC}=1^{--}$:

$$|i\rangle = \frac{1}{\sqrt{2}}\left[|K^0(\vec{p})\rangle|\overline{K}^0(-\vec{p})\rangle - |\overline{K}^0(\vec{p})\rangle|K^0(-\vec{p})\rangle\right]$$

$$= \frac{N}{\sqrt{2}}\left[|K_S(\vec{p})\rangle|K_L(-\vec{p})\rangle - |K_L(\vec{p})\rangle|K_S(-\vec{p})\rangle\right]$$

where $N = (1+|\varepsilon|^2)/(1-\varepsilon^2) \cong 1$ is a normalization factor.

───────────────
*The KLOE collaboration:
A. Aloisio, F. Ambrosino, A. Antonelli, M. Antonelli, C. Bacci, G. Bencivenni, S. Bertolucci, C. Bini, C. Bloise, V. Bocci, F. Bossi, P. Branchini, S. A. Bulychjov, R. Caloi, P. Campana, G. Capon, T. Capussela, G. Carboni, G. Cataldi, F. Ceradini, F. Cervelli, F. Cevenini, F. Chiefari, P. Ciambrone, S. Conetti, E. De Lucia, P. De Simone, G. De Zorzi, S. Dell'Agnello, A. Denig, A. Di Domenico, C. Di Donato, S. Di Falco, B. Di Micco, A. Doria, M. Dreucci, O. Erriquez, A. Farilla, G. Felici, A. Ferrari, M.L. Ferrer, G. Finocchiaro, C. Forti, P. Franzini, C. Gatti, P. Gauzzi, S. Giovannella, E. Gorini, E. Graziani, M. Incagli, W. Kluge, V. Kulikov, F. Lacava, G. Lanfranchi, J. Lee-Franzini, D. Leone, F. Lu, M. Martemianov, M. Matsyuk, W. Mei, L. Merola, R. Messi, S. Miscetti, M. Moulson, S. Müller, F. Murtas, M. Napolitano, A. Nedosekin, F. Nguyen, M. Palutan, E. Pasqualucci, L. Passalacqua, A. Passeri, V. Patera, F. Perfetto, E. Petrolo, L. Pontecorvo, M. Primavera, F. Ruggieri, P. Santangelo, E. Santovetti, G. Saracino, R.D. Schamberger, B. Sciascia, A. Sciubba, F. Scuri, I. Sfiligoi, A. Sibidanov, T. Spadaro, E. Spiriti, M. Testa, L. Tortora, P. Valente, B. Valeriani, G. Venanzoni, S. Veneziano, A. Ventura, S. Ventura, R. Versaci, I. Villella, G. Xu.

#E-mail: antonio.didomenico@roma1.infn.it

As the φ meson is produced almost at rest the kaon momentum is 110 MeV/c and the decay length is 6 mm for $K_S$ and 3.5 m for $K_L$.

The double differential time distribution into final states $f_1$ and $f_2$ at times $t_1$ and $t_2$ is given by:

$$I(f_1,t_1;f_2,t_2) = C_{12}\left\{|\eta_1|^2 e^{-\Gamma_L t_1 - \Gamma_S t_2} + |\eta_2|^2 e^{-\Gamma_S t_1 - \Gamma_L t_2} - 2|\eta_1||\eta_2|e^{-(\Gamma_S+\Gamma_L)(t_1+t_2)/2}\cos[\Delta m(t_1-t_2)+\phi_2-\phi_1]\right\} \quad (1)$$

where $t_1$ ($t_2$) is the time of one (the other) kaon decay into $f_1$ ($f_2$) final state and

$$\eta_i = |\eta_i|e^{i\phi_i} = \langle f_i|K_L\rangle/\langle f_i|K_S\rangle$$

$$C_{12} = \frac{N^2}{2}|\langle f_1|K_S\rangle\langle f_2|K_S\rangle|^2$$

Integrating in $d(t_1+t_2)$ one gets the time difference ($\Delta t = t_1 - t_2$) distribution, sometimes simpler to manipulate:

$$I(f_1,f_2;\Delta t \geq 0) = \frac{C_{12}}{\Gamma_S+\Gamma_L}\left\{|\eta_1|^2 e^{-\Gamma_L \Delta t} + |\eta_2|^2 e^{-\Gamma_S \Delta t} - 2|\eta_1||\eta_2|e^{-(\Gamma_S+\Gamma_L)\Delta t/2}\cos(\Delta m \Delta t + \phi_2 - \phi_1)\right\} \quad (2)$$

valid for $\Delta t \geq 0$, while for $\Delta t < 0$ the substitutions $\Delta t \to |\Delta t|$ and $1 \leftrightarrow 2$ have to be applied. In both (1) and (2) a time interference term is present, characteristic at a φ-factory. As this term oscillates with a period $2\pi/\Delta m \sim 13\tau_S$ but decreases exponentially with a characteristic time $\sim 2\tau_S$, a good decay vertex resolution, i.e. a resolution on kaon decay times $\leq 1 \tau_S$, is crucial for interferometry studies.

From (1) and (2) for different final states ($f_i = \pi^+\pi^-$, $\pi^0\pi^0$, $\pi l\nu$, $\pi^+\pi^-\pi^0$, $3\pi^0$, $\pi^+\pi^-\gamma$, etc.) the kinematical quantities $\Gamma_S$, $\Gamma_L$, $\Delta m$ and module and phase of the $\eta_i$ parameters can be measured. In particular phases $\arg(\eta_i) \equiv \phi_i$ can be measured uniquely for the presence of the interference term. In table 1 a list of the main measurable quantities and the parameters that can be extracted from them is shown; $\varepsilon_K$ and $\delta_K$ are the CP and CPT violation parameters in the kaon mass matrix, respectively; $\varepsilon'/\varepsilon$ is the usual direct CP violation parameter, while a,b,c and d are defined through the semileptonic decay amplitudes as follows:

$$\langle \pi^-\ell^+\nu|K^0\rangle = a+b \qquad \langle \pi^+\ell^-\overline{\nu}|K^0\rangle = c+d$$

$$\langle \pi^+\ell^-\overline{\nu}|\overline{K}^0\rangle = a^*-b^* \qquad \langle \pi^-\ell^+\nu|\overline{K}^0\rangle = c^*-d^*$$

The KLOE interferometry program [2] assumes an integrated luminosity of about 10 fb$^{-1}$, corresponding, at the design luminosity $L = 5\times 10^{32}$ cm$^{-2}$ s$^{-1}$, to about two years of running.

Table 1: List of the main measurable quantities and the corresponding parameters that can be extracted from them.

| Mode | Measured quantity | Parameter |
|---|---|---|
| $\phi \to K_S K_L \to \pi^+\pi^- \, \pi^+\pi^-$ | $I(\pi^+\pi^-,\pi^+\pi^-;\Delta t)$ | $\Delta m$, $\Gamma_S$, $\Gamma_L$ |
| $\phi \to K_S K_L \to \pi^+\pi^- \, \pi^0\pi^0$ | $A(\Delta t) = \dfrac{I(\pi^+\pi^-,\pi^0\pi^0;\Delta t > 0) - I(\pi^+\pi^-,\pi^0\pi^0;\Delta t < 0)}{I(\pi^+\pi^-,\pi^0\pi^0;\Delta t > 0) + I(\pi^+\pi^-,\pi^0\pi^0;\Delta t < 0)}$ | $\Re\left(\dfrac{\varepsilon'}{\varepsilon}\right)$, $\Im\left(\dfrac{\varepsilon'}{\varepsilon}\right)$ |
| $\phi \to K_S K_L \to \pi\ell\nu \, \pi\ell\nu$ | $A_{CPT}(\Delta t) = \dfrac{I(\pi^-e^+\nu,\pi^+e^-\bar\nu;\Delta t > 0) - I(\pi^-e^+\nu,\pi^+e^-\bar\nu;\Delta t < 0)}{I(\pi^-e^+\nu,\pi^+e^-\bar\nu;\Delta t > 0) + I(\pi^-e^+\nu,\pi^+e^-\bar\nu;\Delta t < 0)}$ | $\Re\delta_K - \Re\left(\dfrac{d^*}{a}\right)$ $\Im\delta_K + \Im\left(\dfrac{c^*}{a}\right)$ |
| $\phi \to K_S K_L \to \pi\pi \, \pi\ell\nu$ | $A(\Delta t) = \dfrac{I(\pi^-e^+\nu,\pi^+\pi^-;\Delta t) - I(\pi^+e^-\bar\nu,\pi^+\pi^-;\Delta t)}{I(\pi^-e^+\nu,\pi^+\pi^-;\Delta t) + I(\pi^+e^-\bar\nu,\pi^+\pi^-;\Delta t)}$ | $A_L = 2(\Re\varepsilon_K - \Re\delta_K + \Re b/a + \Re d^*/a)$ $\phi_{\pi\pi}$ |

At present KLOE has collected a total integrated luminosity of ~480 pb$^{-1}$, with a maximum luminosity $L \sim 8\times 10^{31}$ cm$^{-2}$ s$^{-1}$. After the last data taking period DAΦNE has been upgraded, and for year 2004 an increase in luminosity of at least a factor two is expected [1].

## MEASUREMENT OF $\phi \to K_S K_L \to \pi^+\pi^- \, \pi^+\pi^-$

In case of decays of both kaons into $\pi^+\pi^-$ eq.(2) becomes:

$$I(\pi^+\pi^-,\pi^+\pi^-;|\Delta t|) \propto e^{-\Gamma_L|\Delta t|} + e^{-\Gamma_S|\Delta t|} - 2e^{-(\Gamma_S+\Gamma_L)|\Delta t|/2}\cos(\Delta m|\Delta t|) \quad (3)$$

Apart from an overall scale factor, distribution (3) depends only on kinematical quantities [3], and it is shown in Fig.1 with the characteristic lack of events at $\Delta t=0$. In fact a destructive quantum interference prevents the two kaons from decaying into the same final state at the same time. In the figure it is also shown how the distribution changes (dashed and dotted plots) for a $\Delta m$ variation of $\pm 10\%$, corresponding to a change in height and position of the maximum of the distribution.

The experimental resolution on $\Delta t$ in the case of $\pi^+\pi^-$ decays can be improved exploiting the good momentum resolution and the closed kinematics of the event, obtaining with a Monte Carlo (MC) simulation the value $\sigma_{\Delta t} \sim 0.9\ \tau_S$. By fitting the measured distribution including resolution and efficiency effects obtained with MC, and also coherent regeneration effects (see next paragraph), keeping fixed $\Gamma_S$ and $\Gamma_L$ to the PDG [4] values, $\Delta m$ can be evaluated.

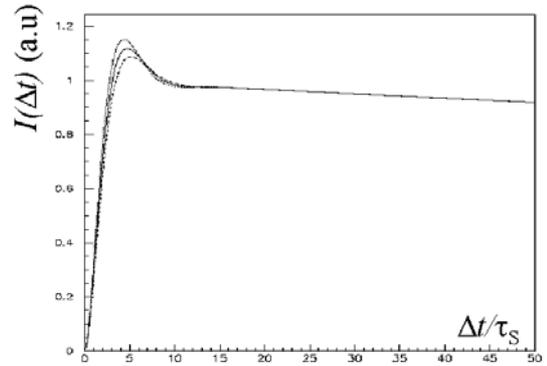

Fig.1: Time difference distribution in the case $f_1=f_2=\pi^+\pi^-$; the dependence on $\Delta m$ variation of $\pm 10\%$ (dashed and dotted plots) is also shown.

The measured distribution obtained with 340 pb$^{-1}$ of data collected in 2001 and 2002 is shown in Fig.2. The KLOE preliminary result is $\Delta m = (5.64 \pm 0.37) \times 10^{-11}$ h/2π s$^{-1}$. The resulting curve of the fit is shown superimposed to data in Fig.2.

This result has to be compared with the more precise PDG value $(5.301 \pm 0.016) \times 10^{-11}$ h/2π s$^{-1}$, showing a correct central value of the KLOE measurement.

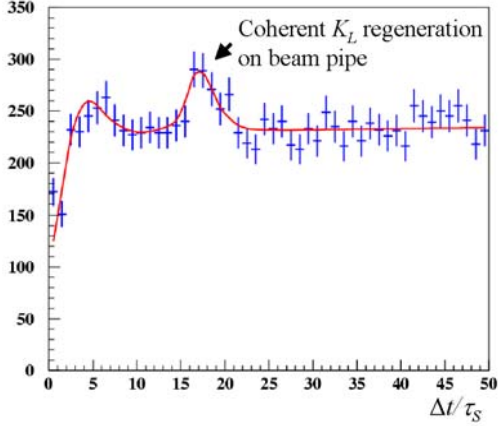

Fig.2: Measured I($\pi^+\pi^-,\pi^+\pi^-;|\Delta t|$) distribution.

It is worth noting that at DAΦNE-2 assuming 500 fb$^{-1}$ of data, after scaling KLOE result for statistics, the current uncertainty on $\Delta m$ could be improved to $\delta\Delta m \sim 0.009 \times 10^{-11}$ h/2π s$^{-1}$.

*Measurement of decoherence parameter*

Quantum theory used to derive distribution (3) could be in principle modified multiplying the interference term for a (1-ζ) factor, where ζ is the decoherence parameter:

$$I(\pi^+\pi^-,\pi^+\pi^-;|\Delta t|) \propto e^{-\Gamma_L|\Delta t|} + e^{-\Gamma_S|\Delta t|}$$
$$- 2 \cdot (1-\zeta) \cdot e^{-(\Gamma_S+\Gamma_L)|\Delta t|/2} \cos(\Delta m|\Delta t|) \quad (4)$$

The case ζ=0 corresponds to the usual "orthodox" quantum theory, while for ζ=1 the case of Furry's hypothesis of spontaneous factorization of wave function (no interference) is obtained [5]. Different ζ values correspond to intermediate situations between these two.

However, as pointed out in Ref. [6], the definition of ζ depends on the basis in which is written the initial state (obviously the "orthodox" quantum theory does not depend on the basis choice). In general for a generic basis $\{K_\alpha,K_\beta\}$, distribution (1) can be modified as follows:

$$I(f_1,t_1;f_2,t_2) = \tfrac{1}{2}\Big[|\langle f_1|K_\alpha(t_1)\rangle\langle f_2|K_\beta(t_2)\rangle|^2$$
$$+ |\langle f_1|K_\beta(t_1)\rangle\langle f_2|K_\alpha(t_2)\rangle|^2 - 2 \cdot (1-\zeta_{K_\alpha,K_\beta}) \cdot$$
$$\cdot \Re\big(\langle f_1|K_\beta(t_1)\rangle\langle f_2|K_\alpha(t_2)\rangle$$
$$\langle f_1|K_\alpha(t_1)\rangle^*\langle f_2|K_\beta(t_2)\rangle^*\big)\Big]$$
$$(5)$$

defining the basis dependent decoherence parameter $\zeta_{K\alpha,K\beta}$.

In Ref.[6] CPLEAR data [7] are re-analysed and the following results are obtained:

$$\zeta_{K_S,K_L} = 0.13 \pm 0.16$$
$$\zeta_{K^0,\overline{K}^0} = 0.4 \pm 0.7 \quad (6)$$

The KLOE preliminary results obtained from a fit to the distribution shown in Fig.2 ($\Delta m$ fixed) are:

$$\zeta_{K_S,K_L} = 0.12 \pm 0.08$$
$$\zeta_{K^0,\overline{K}^0} = (0.8 \pm 0.5) \times 10^{-5} \quad (7)$$

with an improvement on both measurements. In particular the result in the case of the flavour basis benefits of very large cancellations (occurring only in the $\pi^+\pi^-$ final state case, contrary to the CPLEAR case) between the interference term and the other two terms in (5), resulting in a function very sensitive to ζ deviations from zero.

## C-EVEN BACKGROUND

In principle, the decay of scalar mesons into a neutral kaon pair could be a dangerous background for interferometry measurements, because in this case the kaon pair is in a C-even state:

$$|i\rangle = \frac{1}{\sqrt{2}}\{K^0\overline{K}^0 + \overline{K}^0 K^0\} \cong \frac{1}{\sqrt{2}}\{K_S K_S - K_L K_L\}$$

Recent evaluations of BR[$\phi \to (f_0+a_0)\gamma \to K^0\underline{K}^0\gamma$] report a value in the range (1÷4) ×10$^{-8}$ [8,9] (the most recent one is based on KLOE data [10,11] on scalar mesons). Given this value of the branching ratio the impact of the C-even background on interferometry measurements is negligible. In any case it can be still largely reduced exploiting:

- the different spatial behaviour of kaon decays in the C-even and C-odd cases;
- the detection of the low energy radiated photon;
- the measurement of one kaon momentum and the kinematics constraints for the C-even background.

In case one or both kaons decay into $\pi^+\pi^-$, the last point can be very effective in rejecting the background, thanks to the good KLOE momentum resolution: $\sigma(P_K)\sim 1.8$ MeV/c.

In fact due to kinematics constraints (dashed band in Fig.4), an acceptance 3σ-cut around the expected momentum $P_K=110$ MeV/c for the C-odd signal will select a kaon pair invariant mass $M_{KK} > 1017$ MeV. As the expected $M_{KK}$ spectrum for the C-even background has a negligible contribution in that region [8,9], the previous cut will reject the vast majority of background without loosing the signal.

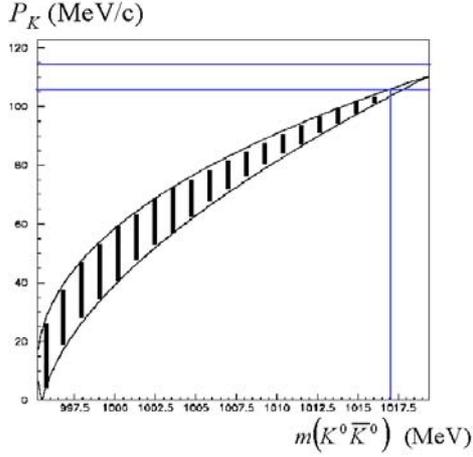

Fig.3: The dashed band in the {$P_K$, $M_{KK}$} plane representing the kinematics constraints for the C-even background; a 3σ-cut around $P_K$=110 MeV/c and the corresponding bound on $M_{KK}$ is also shown.

## REGENERATORS

In general at a φ-factory regenerators are always present and have to be kept far away from the interaction point (IP) in order to not distort the shape of signal distributions in the interference region. At KLOE there are essentially three regenerators, as sketched in Fig.4:

- a spherical 500 μm thick beam pipe made of an Al-Be alloy at a distance of 10 cm from the IP;
- a cylindrical 50 μm thick pure Be foil at a distance of 4.4 cm from the IP;
- the drift chamber (DC) inner wall made of carbon fiber (675 μm thickness) and two Al foils (100 μm thickness each) at distance of 25 cm from the IP.

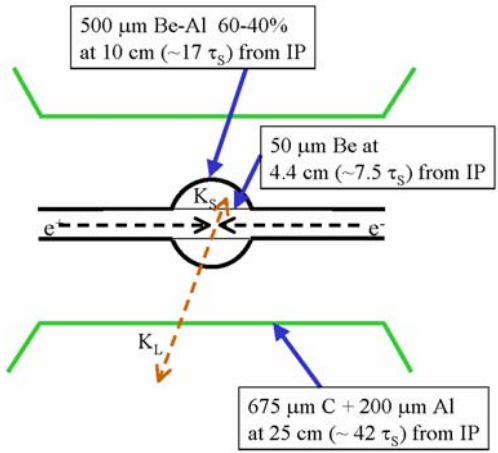

Fig.4: Sketch of IP and regenerators at KLOE

In Fig.5 the measured distribution of the distance from the beam line of the reconstructed $K_{S,L} \to \pi^+\pi^-$ decay vertices is shown. The three peaks due to the incoherent regeneration $K_L \to K_S \to \pi^+\pi^-$ on the regenerators are clearly visible. However incoherent regeneration occurs at large angle with respect to the initial $K_L$ direction and can be easily recognised and rejected.

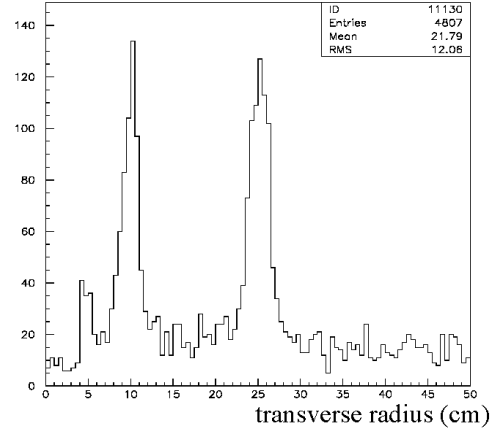

Fig.5: Distribution of the distance from the beam line of the reconstructed $K_{S,L} \to \pi^+\pi^-$ decay vertices.

On the other hand coherent regeneration occurs strictly in the forward direction, interferes with the signal and cannot be distinguished from it. For this reason it is crucial to have regenerators far away from the IP, as in the case of Fig.2, where the irreducible coherent regeneration contribution is far from the interference region at Δt~0.

The intensity of decays into charged pions of a $K_L$ beam, encountering a regenerator at a proper time t=0, is:

$$I(\pi^+\pi^-, t) = \left|\langle\pi^+\pi^-|K_S\rangle\right|^2 \left[|\eta_\pm|^2 e^{-\Gamma_L t} + |\rho_{coh}|^2 e^{-\Gamma_S t} \right.$$
$$\left. + 2|\eta_\pm||\rho_{coh}|e^{-(\Gamma_S+\Gamma_L)t/2}\cos(\Delta m t + \phi_{coh} - \phi_\pm)\right]$$
(8)

where $|\rho_{coh}|$ and $\phi_{coh}$ are modulus and phase of the coherent regeneration parameter. In case of thin regenerators (as for KLOE) the interference term in (8) is not negligible. The measured proper time distribution of $K_{S,L} \to \pi^+\pi^-$ decay vertices behind the DC inner wall is shown as an example in Fig.6, after rejection and subtraction of the residual incoherent background. A fit with distribution (8) including resolution effects gives the following preliminary result for the coherent regeneration parameter in the DC wall:

$$|\rho_{coh}| = (16.6 \pm 6.6) \times 10^{-4}$$
$$\phi_{coh} = -1.19 \pm 0.27 \text{ rad}$$
(9)

The black curve in Fig.7 represents the fit result; the red curve shows, for comparison, how the incoherent background shape (i.e. without interference term) would appear, confirming that its behaviour cannot be confused with the coherent regeneration one.

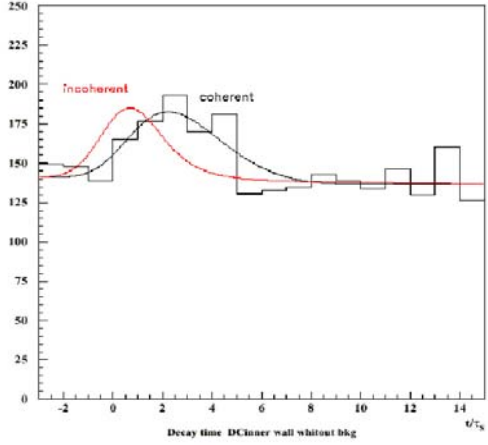

Fig.6: The measured proper time distribution of $K_{S,L} \to \pi^+\pi^-$ decay vertices behind the DC inner wall.

## FUTURE

At the moment the KLOE analysis tools for interferometry are being refined. The interferometry program needs to start at least half of the expected 10 fb$^{-1}$ of data. As an example the estimated uncertainties on CP and CPT parameters expected with 10 fb$^{-1}$ are:

$$\Delta\left[\Re\left(\frac{\varepsilon'}{\varepsilon}\right)\right] \sim \Delta\left[\Re\delta - \Re\left(\frac{d^*}{a}\right)\right] \sim 5 \times 10^{-4}$$

$$\Delta\left[\Im\left(\frac{\varepsilon'}{\varepsilon}\right)\right] \sim \Delta\left[\Im\delta - \Im\left(\frac{d^*}{a}\right)\right] \sim 10^{-2}$$

As the dependence of distribution (2) on the imaginary parts of the parameters is concentrated in the interference term, the imaginary parts are less precisely determined than the real ones. In fact the interference term lasts only few $\tau_S$ and its measurement does not benefit of the whole statistics of the distribution; moreover the presence of rapid oscillations makes the determination affected by vertex resolution effects.

At DAΦNE-2, assuming 500 fb$^{-1}$ of data, the knowledge of the kinematical quantities $\Gamma_S$, $\Gamma_L$ and $\Delta m$ should be improved, as previously mentioned. Also the measurement of imaginary parts of CP and CPT violation parameters would benefit of larger statistics and could be measured at a precision level comparable with the present one on real parts. Moreover interferometry with very rare final states could be feasible or improved; finally several proposed tests of quantum mechanics at a φ-factory [12,13] – almost all requiring very high statistics – could be performed.